\documentclass[11pt, a4paper, logo, copyright, nonumbering]{kwaipilot}
\usepackage{dblfloatfix}
\usepackage{ulem}
\usepackage{caption}
\usepackage{xspace}
\usepackage{pifont} %
\usepackage{multirow}
\usepackage{tcolorbox}
\usepackage{xltabular}
\usepackage{longtable}
\usepackage{hyperref}
\usepackage{tabularx}
\usepackage{booktabs}
\usepackage{array}
\usepackage{graphicx}
\usepackage{graphicx}
\usepackage{subcaption}
\usepackage{float}
\usepackage{amsfonts}
\usepackage{amsmath}
\usepackage{amssymb}
\usepackage{lineno}
\usepackage{multirow}
\usepackage{adjustbox}
\usepackage{bbm}
\usepackage{multirow}
\usepackage{booktabs}
\usepackage{array}
\usepackage{ragged2e}

\usepackage{booktabs}
\usepackage{array}
\usepackage{multirow}
\usepackage{makecell}
\usepackage{ragged2e}

\newcolumntype{C}[1]{>{\centering\arraybackslash}m{#1}}
\newcolumntype{L}[1]{>{\RaggedRight\arraybackslash}m{#1}}

\usepackage{longtable}
\usepackage{array}
\usepackage{ragged2e}
\usepackage{needspace}

\newcolumntype{L}[1]{>{\RaggedRight\arraybackslash}p{#1}}
\usepackage{underscore}

\usepackage[bottom]{footmisc}

\usepackage{CJKutf8}
\usepackage{setspace}

\usepackage{dsfont}
\usepackage{array} %
\usepackage{tabularx} %
\usepackage{xcolor} %
\usepackage{tabularx}
\usepackage{booktabs}

\usepackage{lipsum}  %
\usepackage{multicol} %

\usepackage{indentfirst}
\usepackage[sort&compress,numbers]{natbib}
\newcolumntype{L}[1]{>{\raggedright\arraybackslash}p{#1}}

\usepackage{geometry}
\usepackage{makecell}
\makeatletter
\def\@BTrule[#1]{%
  \ifx\longtable\undefined
    \let\@BTswitch\@BTnormal
  \else\ifx\hline\LT@hline
    \nobreak
    \let\@BTswitch\@BLTrule
  \else
     \let\@BTswitch\@BTnormal
  \fi\fi
  \global\@thisrulewidth=#1\relax
  \ifnum\@thisruleclass=\tw@\vskip\@aboverulesep\else
  \ifnum\@lastruleclass=\z@\vskip\@aboverulesep\else
  \ifnum\@lastruleclass=\@ne\vskip\doublerulesep\fi\fi\fi
  \@BTswitch}
\makeatother

\addto\extrasenglish{
}

 {\begin{list}{}%
         {\setlength{\leftmargin}{#1}}%
         \item[]%
 }
 {\end{list}}

\reportnumber{001} %

\renewcommand{\today}{}

\title{\centering KAT-Coder-V2.5 Technical Report}

\author[*]{
KwaiKAT Team
}

\begin{document}

\begin{abstract}
\small{We present {KAT-Coder-V2.5}, a coding-focused agentic model trained to act autonomously inside real, executable repositories rather than as a single-turn code generator. Its capability is bottlenecked less by model scale than by the scarcity of reproducible environments, verifiable rewards, and high-value trajectories, which we address with an end-to-end agentic post-training framework. {AutoBuilder} reconstructs multilingual repositories into sandboxed environments with fail-to-pass and pass-to-pass verification at scale, from which we regenerate self-contained task specifications, recover near-miss trajectories, and distill supervision through process-aware filtering, while {KwaiClawEnv} synthesizes large-scale tool-use trajectories from executable services and real task seeds. We further scale reinforcement learning with harness randomization, a reliability-hardened sandbox, an asymmetric actor--critic PPO with hindsight-augmented value estimation, and a harness-oriented reward framework, and unify SWE, Agent-Claw, and WebCoding experts via {Multi-Teacher On-Policy Distillation}. Across six software-engineering and agentic benchmarks, KAT-Coder-V2.5 delivers the best agentic tool-use result on PinchBench and ranks second only to the frontier Opus 4.8 on repository-level software engineering. Our service is available at \url{https://streamlake.com/product/kat-coder}.}
\end{abstract}

\maketitle

\begin{figure}[htbp]
    \centering
    \includegraphics[width=0.91\textwidth]{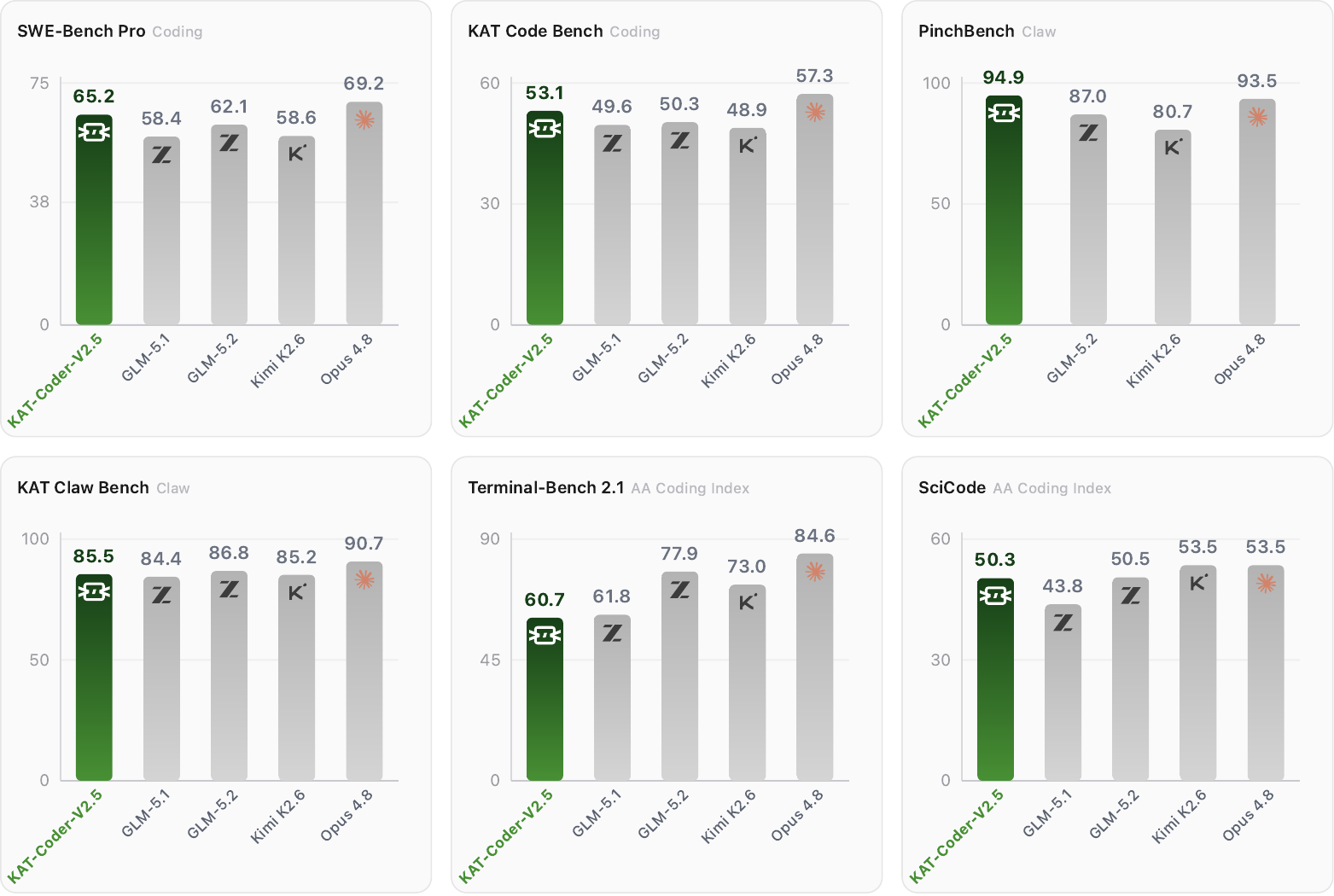}
    \caption{\small{Results of KAT-Coder-V2.5 on various SWE and agent benchmarks.}}
    \label{fig:kat-coder-v25-benchmarks}
\end{figure}

\section{Introduction}
Coding models~\cite{katcoderv2} have moved beyond passive code completion toward autonomous agents that understand requirements, navigate repositories, edit files, run tests, and iteratively repair their own failures~\cite{jimenez2024swebench,deng2025swebenchproaiagents}. This shift redefines what capability means: a strong coding agent must combine repository understanding, tool use, long-horizon planning, engineering discipline, and verification-driven problem solving, not merely generate syntactically correct snippets. Requirements are stated in natural language, relevant logic is scattered across many files, and the agent must search, localize, design, edit, verify, and recover under limited context.

We argue that the primary obstacle to building such agents is not model scale but the training infrastructure around it, which manifests as three concrete challenges.

\noindent\textbf{Scalable executable environments.} Real repositories differ substantially in dependencies, build systems, testing frameworks, and runtime assumptions. Producing environments that are reproducible, executable, and truly verifiable is difficult at scale, and naive construction silently admits environments that never run the intended tests~\cite{yang2025swe,badertdinov2026swe}.

\noindent\textbf{Trajectory quality beyond final rewards.} Raw issues and pull-request descriptions are often incomplete, ambiguous, or inconsistent with the merged change. Worse, filtering trajectories by final test success alone is misleading: some passing trajectories rely on hard-coding, mechanism bypassing, or test-oriented shortcuts, while some failed trajectories still contain valuable search, localization, and repair behavior.

\noindent\textbf{Stable long-horizon RL.} Long-horizon agentic reinforcement learning suffers from sparse rewards, unstable environment feedback, coarse credit assignment, overfitting to a fixed harness, and difficulty in fusing capabilities across specialized experts.

To address these challenges, we build KAT-Coder-V2.5 around a systematic post-training framework that treats agentic capability as a systems problem, not merely a matter of scaling data or parameters. For software-engineering tasks, we propose {AutoBuilder}, an agent-driven pipeline that reconstructs multilingual, reproducible, executable repository environments and formulates each SWE sample as a verifiable task defined by a task description, an execution environment, and a verifier with fail-to-pass and pass-to-pass tests. From golden and test patches, we regenerate structured, self-contained task descriptions so the model trains on clear executable requirements rather than noisy raw issue text, and we introduce {process-aware trajectory construction} that evaluates exploration, localization, design, editing, verification, and recovery behavior, removing passing-but-undesirable trajectories and recovering informative near-miss failures into high-quality samples. For general agentic capabilities, we develop {KwaiClawEnv}, a scalable environment synthesis framework organized around Service, Task, and Eval layers that produces large-scale tool-use trajectories over heterogeneous tools, multi-service workflows, and long-horizon reasoning.

During reinforcement learning, we treat the harness itself as part of the training distribution, introducing both white-box and black-box harnesses to reduce overfitting to a single tool protocol, context format, or control-flow pattern. We couple this with a reliability-hardened sandbox that reduces reward corruption from environment errors, and combine an asymmetric actor--critic PPO with a harness-oriented reward framework to deliver stable, fine-grained signals for long-horizon tasks. Finally, we train a set of specialized experts spanning agentic software engineering, agentic tool use (Claw), terminal use, web coding, and general knowledge, and fuse them through {Multi-Teacher On-Policy Distillation} on trajectories sampled from the student's own policy, alleviating the see-saw effect common in multi-expert merging. In this report we focus on the software-engineering and Claw experts, which best illustrate our environment-construction and trajectory-filtering methodology, while the remaining experts follow the same recipe.
Our main contributions are summarized as follows:
\begin{itemize}
    \item We build a \textbf{verifiable SWE training infrastructure} with AutoBuilder, which constructs reproducible repository environments and grounds tasks with executable fail-to-pass and pass-to-pass verification signals.

    \item We introduce a \textbf{process-aware trajectory pipeline} that filters agent behavior beyond final pass rates and recovers near-miss failures through hint-free reasoning regeneration.

    \item We establish a \textbf{unified agentic post-training framework} that combines KwaiClawEnv, harness randomization, a reliability-hardened sandbox, asymmetric PPO, reward modeling, and multi-teacher on-policy distillation.

    \item We propose two internal benchmarks, \textbf{KAT Code Bench} and \textbf{KAT Claw Bench}, to evaluate coding agents under realistic engineering tasks and tool-use workflows.
\end{itemize}

Evaluated under a unified Claude Code harness across six benchmarks, KAT-Coder-V2.5 shows strong performance in agentic tool use and repository-level software engineering, achieving the top result on PinchBench and ranking second on both SWE-Bench Pro and KAT Code Bench among the evaluated models.

\section{Agentic Software-Engineering Capabilities}

\begin{figure}[tbp]
\centering
\includegraphics[width=1.0\linewidth]{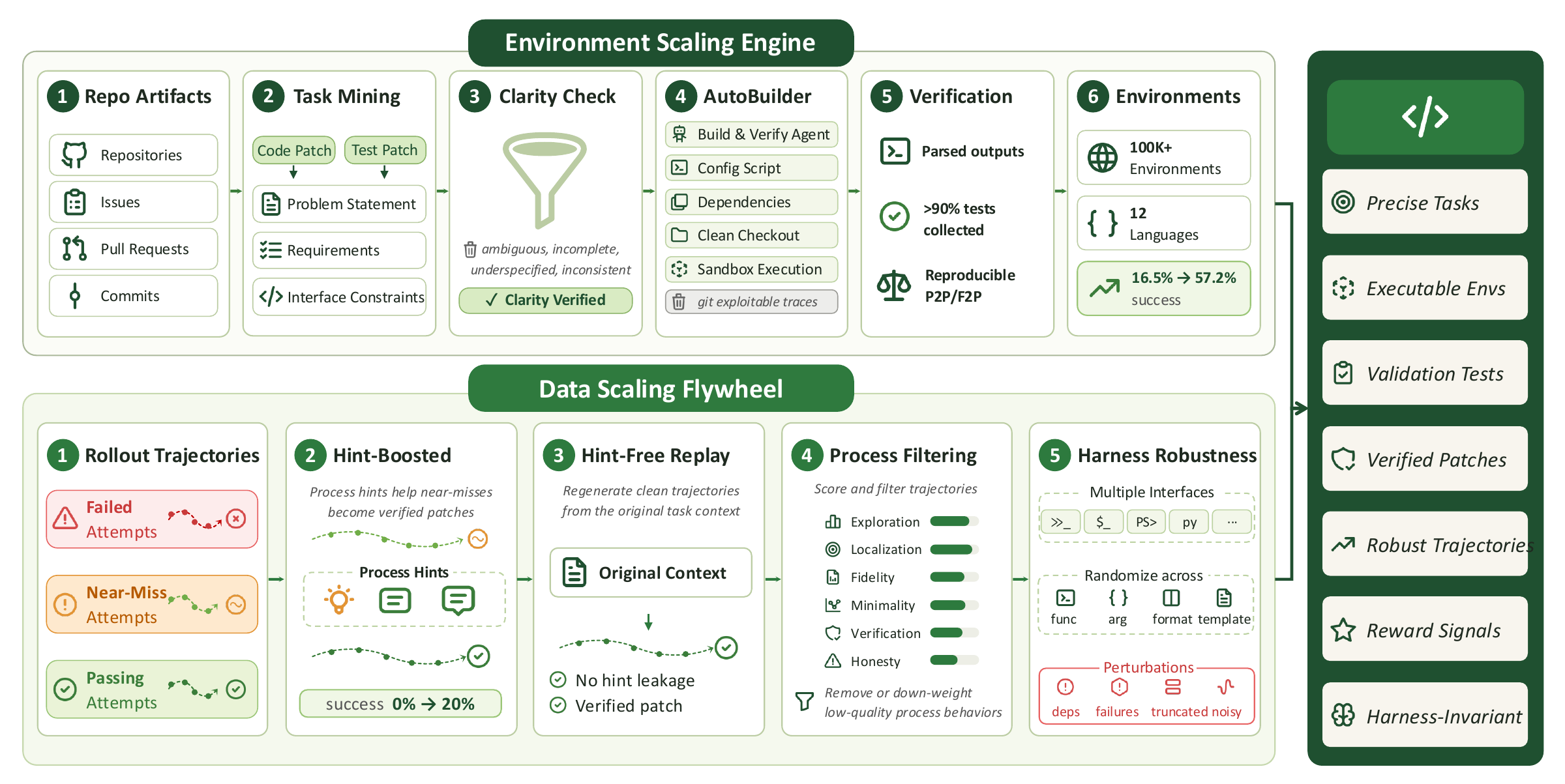}
\caption{The agentic software-engineering data pipelines.}
\label{fig:swe-pipeline}
\end{figure}

Agentic software engineering requires tasks that are precisely specified, executable, and objectively verifiable in reproducible environments. We scale agentic software-engineering capabilities along two complementary axes. \textit{Environment Scaling} constructs large-scale executable and verifiable task environments from real repositories, while \textit{Data Scaling} turns rollout trajectories into higher-value training signals through failure recovery, process-aware filtering, and distributional generalization.

\subsection{Environment Scaling Engine}
\label{sec:swe-environment-scaling}

Scaling verifiable software-engineering tasks from real-world repositories poses two central challenges. First, raw issue and PR descriptions are often ambiguous, incomplete, project-specific, or misaligned with the final merged changes, making them unreliable as task specifications. Second, repository environments are brittle to reconstruct at scale, and naive validation based on command exit codes or surface-level logs can falsely accept environments that never execute the intended tests. We address these challenges through verifiable task mining and automated environment construction.

\paragraph{Verifiable Task Mining.}
\label{sec:swe-task-mining}

A {verifiable task} is defined as a triplet consisting of a precise task description, an executable repository environment, and a set of validation tests used to determine correctness. Given such a task, an agent starts from the initial repository state and produces a patch; the patch is considered correct if and only if it passes all validation tests.

We mine tasks primarily from real pull requests and commits, where the merged code change provides a {golden patch} and the accompanying test change provides a {test patch}~\cite{jimenez2024swebench}. However, raw issue and PR text is often unreliable: it may be vague, incomplete, inconsistent with the final merged change, or heavily dependent on project-specific context. Therefore, we regenerate structured task descriptions from the underlying verifiable artifacts. Each description contains three components: a {problem statement} that characterizes the bug or missing functionality, grounded mainly in the golden patch; {requirements} that specify the expected behavior, derived primarily from the test patch; and {interface constraints} that define the APIs, variables, data structures, and compatibility assumptions the solution must respect, inferred from both patches. We then apply a clarity check to remove samples whose descriptions are ambiguous, incomplete, underspecified, or internally inconsistent, ensuring that each retained task is precise, self-contained, and aligned with the validation tests.

\paragraph{Verifiable Environment Construction.}
\label{sec:swe-env-scaling}

We introduce \textbf{AutoBuilder}, an agent-driven pipeline for multilingual execution-environment construction. {AutoBuilder} operates as a sandboxed build--verification loop: a build agent analyzes the repository and generates a configuration script that installs dependencies and runs tests from a clean checkout, while a verification agent executes the script in an isolated sandbox and validates the result. Crucially, verification does not rely on command exit codes or superficial log patterns. Instead, it parses structured test-framework outputs and accepts an environment only when more than 90\% of the expected tests are collected and the pass/fail outcomes are reproducible across runs. When verification fails, structured failure information is fed back to the build agent for iterative repair.

For scalability, {AutoBuilder} combines a preconfigured base environment, language- and build-system-specific templates, and a retrievable library of successful configurations distilled into reusable build recipes. These components reduce unnecessary trial-and-error in dependency installation and test invocation, increasing the environment-construction success rate from 16.5\% to 57.2\% and yielding over 100{,}000 verifiable environments across 12 languages.

To keep agent rollouts focused on code reasoning rather than repository setup or environment manipulation, we pre-apply dependency updates and environment-configuration edits from the reference change when they are not part of the intended programming challenge. We also remove git history, commit metadata, and other exploitable traces that could reveal the reference solution, so that agents must solve the task from the provided description, repository state, and executable tests alone.

\subsection{Data Scaling Flywheel}
\label{sec:swe-data-scaling}

Data Scaling turns rollout trajectories into higher-value training signals by recovering informative failures, filtering undesirable behavior, and broadening the deployment distribution observed during training.

\paragraph{Hint-Boosted Rollout Pass Rate}
\label{sec:swe-hint-loop}

Many failed trajectories are near misses: the model localizes the relevant code region but misses a decisive step, such as reading the decisive test assertion, matching an exact schema, reusing an existing mechanism, or continuing diagnosis after an initial failure. Instead of discarding such trajectories, we recover them with a two-stage hint-in-the-loop procedure. First, we inject targeted process-level hints that indicate what to inspect or verify without revealing the solution, which alone raises the pass rate of previously zero-pass tasks to roughly 20\%. Since hinted trajectories may contain information unavailable at inference time, we then fix the verified patch and regenerate a hint-free trajectory from the original task context, retaining only samples that pass verification, contain no hint leakage, and remain consistent with the verified patch and test outcomes. Hints thus act only as temporary scaffolding, while the final training data stays faithful to the original task distribution.

\paragraph{Process-Score-Driven Trajectory Filtering}
\label{sec:swe-traj-filter}

Passing tests is necessary but not sufficient for high-quality supervision: some successful trajectories rely on brittle shortcuts, hard-coded behavior, test tampering, or weak verification. Large-scale replay reveals recurring low-quality patterns, including insufficient exploration, poor localization, poor specification matching, bypassing existing repository mechanisms, and incomplete validation. We therefore combine rule-based gates with heuristic process scoring: the rule-based stage removes invalid, unstable, or exploitative trajectories, while the scoring stage evaluates process dimensions such as exploration, localization, pre-edit reasoning, specification fidelity, adherence to repository conventions, patch minimality, verification quality, recovery behavior, and honesty. Shortcut-based successes are down-weighted or removed, and recoverable near misses are routed back into the recovery pipeline above. These process annotations also provide positive and negative trajectory signals for preference learning, rejection sampling, and process reward modeling.

\paragraph{Harness Rewriting for Robustness}
\label{sec:swe-generalization}

Training only in a fixed harness and clean sandbox can induce interface overfitting and weak recovery under realistic failures. To improve cross-harness generalization, we rewrite tool names, argument conventions, output formats, and prompt templates through randomization while preserving their underlying functionality. Because verification is anchored to structured test outcomes rather than harness-specific traces, the same task can be re-served under multiple equivalent harness configurations, forcing the model to learn harness-invariant problem-solving strategies. On top of this, we inject realistic perturbations—missing or mismatched dependencies, transient command failures, truncated outputs, and noisy logs—so the model learns to continue diagnosis and verification under anomalous execution conditions rather than stopping at the first failure signal.

\section{General Agentic Capabilities}

\begin{figure}[tbp]
\centering
\includegraphics[width=0.95\linewidth]{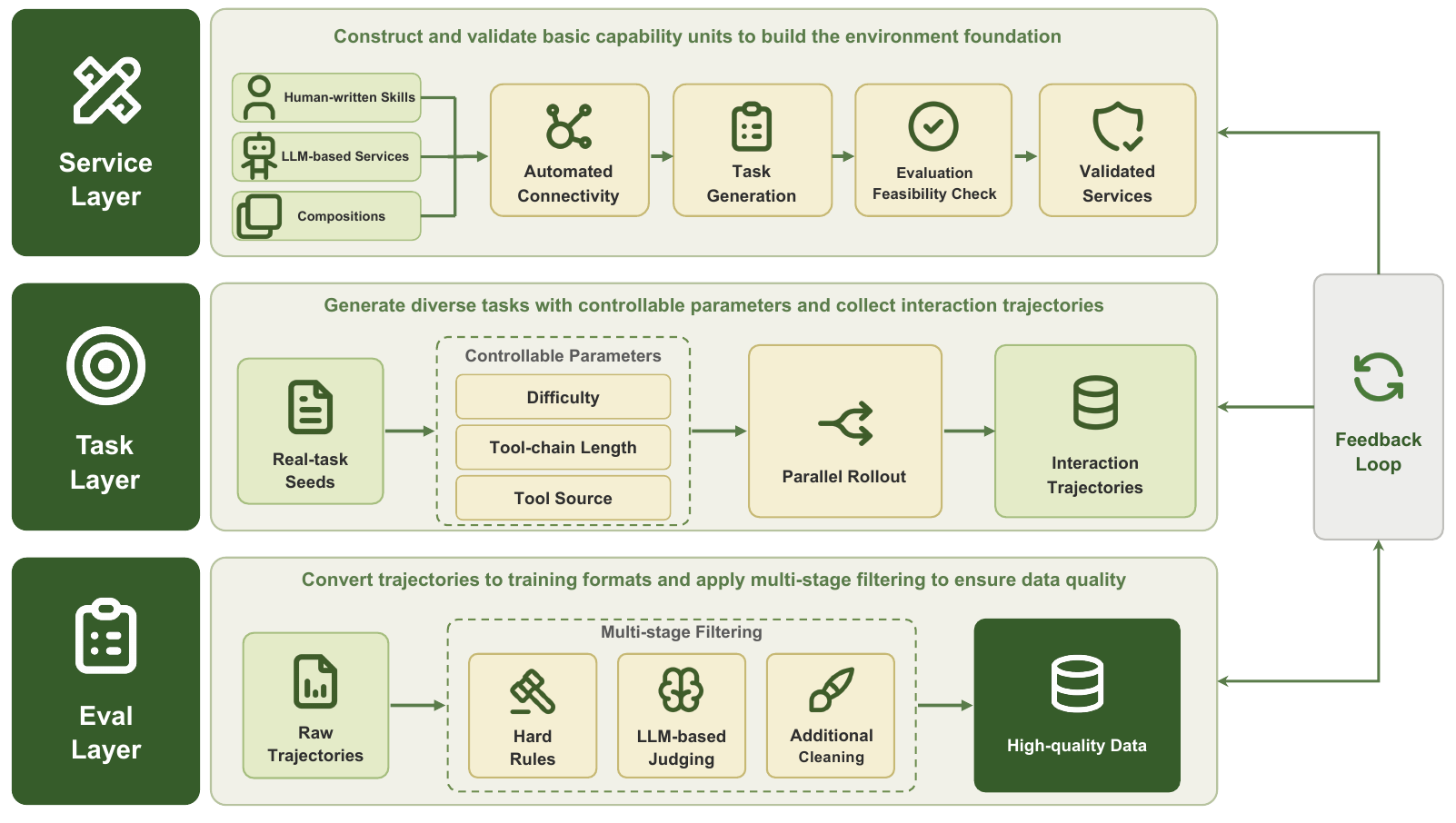}
\caption{Pipeline of KwaiClawEnv.}
\label{fig:claw_pipeline}
\end{figure}

\subsection{Overview of KwaiClawEnv}

To evolve LLMs from passive language processing tools into agents capable of autonomous decision-making and action in real-world settings, models must develop strong multi-environment adaptability and robust tool-use capabilities. Achieving this requires large-scale training in high-quality, diverse, and interactive environments.

Existing approaches face several key limitations: real systems are difficult to access at scale due to permission, security, and stability constraints. LLM-simulated environments often suffer from hallucinations and inconsistent states. And manually built sandboxes, while reliable, are costly and hard to scale. In addition, open-source datasets are often biased toward a limited set of scenarios, leaving long-tail business needs underrepresented and restricting model generalization. Existing environment synthesis methods also tend to be environment-specific, lacking unified modeling and transferability across heterogeneous settings~\cite{zheng2023docarena}.

To address these challenges, we propose KwaiClawEnv, an environment synthesis framework designed around real business needs and Claw-style Agent tasks~\cite{openclaw2026}. KwaiClawEnv adopts a structured generation paradigm based on Skill templates and real Tasks~\cite{skill2026pseudocode}, ensuring semantic consistency, logical completeness, and reliable state transitions while reducing hallucination-related quality issues. The framework supports unified modeling across heterogeneous environments, enabling capability transfer across tools, tasks, and interaction workflows. It also provides controllable task complexity by flexibly adjusting tool-chain length, task steps, and interaction difficulty, supporting scalable data construction from basic tool use to complex multi-step reasoning. Overall, KwaiClawEnv offers a high-quality, cost-effective, and scalable solution for large-scale agent training~\cite{li2026clawenvkit,bai2026clawgym}.

\subsection{Pipeline of KwaiClawEnv}

KwaiClawEnv is organized around three core questions: how to generate data, how to ensure data quality, and how to continuously improve the generation process. To address these challenges, it adopts a three-layer architecture consisting of the \textbf{Service}, \textbf{Task}, and \textbf{Eval} layers, which are connected through a closed feedback loop. The Service layer constructs executable capabilities, the Task layer generates tasks and samples interaction trajectories~\cite{trajectory2026large}, and the Eval layer filters and formats the resulting data while feeding quality signals back to earlier stages for iterative improvement.

\subsubsection{Service Layer: Capability Construction and Filtering}

The Service layer defines the atomic capability units of KwaiClawEnv and provides the executable environment for agent interaction. It supports two categories of sources: human-authored \textbf{Skills} and LLM-generated \textbf{Services}. Based on these atomic Services, the system further constructs composite capabilities by chaining or nesting multiple Services to satisfy real-world task requirements.

Before being admitted into the task-generation pipeline, each Service undergoes automatic validation to ensure executability, interface consistency, and logical correctness. Only Services that pass this validation are retained for downstream task generation and trajectory sampling.

\subsubsection{Task Layer: Task Generation and Trajectory Sampling}

The Task layer instantiates concrete tasks from the available Service pool and executes them to obtain complete interaction trajectories. Starting from real tasks as seeds, the system derives multiple task variants that differ in execution paths, tool combinations, and task constraints, thereby expanding task diversity while preserving real-world relevance.
Task generation is governed by configurable parameters, including task difficulty, tool-chain length, and tool source. During execution, the system performs parallel rollouts and records the full interaction process, including model decisions, tool invocations, tool outputs, and state transitions. These records form end-to-end traceable trajectories from user input to final task completion.

\subsubsection{Eval Layer: Data Processing and Quality Control}

The Eval layer processes and filters the trajectories produced by the Task layer. Raw trajectories are first converted into a unified training format, such as SFT-ready samples, with missing or auxiliary fields populated to ensure consistency across samples. A multi-stage filtering pipeline is then applied to retain only reliable and high-quality trajectories.

The resulting quality signals are propagated back to the Service and Task layers, enabling continuous refinement of subsequent data generation. Through this closed-loop mechanism, KwaiClawEnv can iteratively improve both capability construction and task generation, ultimately producing scalable, diverse, and high-quality agent training data. Evaluation follows trustworthy agent evaluation principles including trajectory-transparent grading and multi-dimensional quality assessment~\cite{ye2026claweval,pinchbench}.

\subsection{Environment Scaling}

To construct large-scale, diverse, and high-quality training data, KwaiClawEnv introduces a two-level scaling mechanism that combines Service-level expansion with Task-level derivation. By jointly scaling executable Services and derived Tasks, the system amplifies a small set of high-quality seeds into tens of thousands of diverse interaction trajectories, enhancing the model's adaptability to long-tail scenarios, heterogeneous tools, and complex workflows.

\subsubsection{Service-level Scaling: From Skills to Executable Environments}

At the Service layer, KwaiClawEnv employs a dual-source generation strategy. The first source leverages standardized Skill definitions from open-source communities such as OpenClaw. The system parses these definitions to extract API specifications, parameter schemas, and usage constraints, then converts them into deployable services equipped with OpenAPI specifications, container configurations, and fixture data. This approach inherits quality signals from curated community Skills and achieves a generation success rate exceeding 90\%.

The second source uses category-guided LLM generation to synthesize new Skill variants for underrepresented domains. Starting from a set of atomic Services, the system further constructs composite environments through service chaining and orchestrated combination. After applying semantic constraints and multi-stage validation, a large pool of high-quality environment compositions is retained, expanding the environment scale by an order of magnitude compared with manual construction.

\subsubsection{Task-level Scaling: Seed Derivation and Complexity Control}

At the Task layer, KwaiClawEnv employs a structured derivation engine to expand real business tasks into large-scale training instances. The process begins with high-quality seed tasks, each defined by a clear objective, tool-use exemplars, and machine-verifiable success criteria. The engine then generates variants through three mechanisms: parameter expansion, constraint augmentation, and tool-chain orchestration.

Task complexity is governed by configurable controls including difficulty ratio, tool-chain length bounds, and tool source composition. In practice, KwaiClawEnv generates millions of candidate tasks and retains over one hundred thousand high-quality instances after multi-stage verification. The resulting trajectories contain an average of 15 tool calls, with the longest exceeding 100 steps, covering scenarios that range from basic single-tool usage to complex long-horizon reasoning.

\subsection{Data Validation and Reliability Assurance}

KwaiClawEnv significantly improves the efficiency of environment construction. However, large-scale automated generation also introduces new quality-control challenges. In early experiments, we observed consistency issues across service generation, task synthesis, and evaluation execution. For instance, generated task files may fail schema validation, tool calls may reference unimplemented endpoints, and synthesized trajectories may conflict with fixture data. While such issues are manageable at a small scale, they can accumulate and propagate as the system scales to hundreds of services and thousands of tasks.

To address these challenges, we design an end-to-end consistency validation framework that covers the full lifecycle from service generation to task execution. The framework follows a progressive three-stage pipeline. First, \textbf{service availability validation} checks endpoint reachability, OpenAPI specification completeness, and inter-service dependency connectivity. Second, \textbf{task generation validation} verifies schema legality, tool-reference correctness, parameter consistency, and the machine-verifiability of success criteria. Third, \textbf{execution environment validation} runs tasks within a containerized sandbox to verify service startup, fixture loading, agent interaction, trajectory integrity, and scoring correctness. Each sample is assigned a unified disposition, namely \textit{repairable failure}, \textit{rejected defect}, or \textit{production-ready output}, enabling targeted remediation.

At the trajectory level, KwaiClawEnv further applies a two-layer filtering strategy. The first layer, \textbf{hard-rule filtering}, removes unsafe, invalid, or hallucinated trajectories through deterministic checks, including tool-blacklist detection, file-existence validation, mandatory tool-coverage verification, and state-consistency enforcement. The second layer, \textbf{LLM-as-Judge evaluation}, scores the remaining trajectories along three dimensions: semantic correctness, execution efficiency, and interaction naturalness. Only trajectories that pass both layers are retained for training.

This framework enables KwaiClawEnv to expand a limited set of atomic services into diverse, high-quality environments and to produce reliable trajectories covering representative scenarios such as coding, writing, and data analysis.

\section{Reinforcement Learning}

\subsection{Harness Scaling}

In agentic RL, if training relies solely on a single fixed harness, the model often learns not "how to solve the task" but "how to solve the task under that particular harness's interface conventions." This implicit coupling manifests as overfitting in three forms:
\begin{itemize}
    \item Format overfitting: The model becomes anchored to a specific action format, and the parsing failure rate rises significantly once the tool-invocation protocol is switched.
    \item Context-structure overfitting: The model depends on the training harness's specific history-concatenation order, and its behavior degrades when the context layout is rearranged.
    \item Control-flow overfitting: The model relies on the reflection timing and stopping conditions provided by the training harness, and fails to plan autonomously in harnesses that lack such explicit scaffolding.
\end{itemize}
Therefore, we introduce diverse harnesses during the RL rollout phase, so that the model's competence consolidates at the level of "task solving" rather than "interface adaptation," thereby achieving cross-harness generalization. In essence, this is a form of domain randomization applied at the environment level.
The key to Harness Scaling lies not in the number of harnesses but in whether the diversity falls along dimensions that are useful for generalization. We systematically construct diversity along the following "axes of variation":
\begin{itemize}
    \item Tool-invocation protocol: structured function-calling, text code-block protocols, tag-based protocols, and others, mitigating format overfitting.
    \item Context-management strategy: full history, sliding window, summary compression, and different observation-truncation strategies, improving the model's robustness in scenarios with incomplete information.
    \item Control-flow complexity: ranging from minimal ReAct to complex control flows with explicit planning and self-reflection, enabling the model to operate under both "guided" and "unguided" conditions.
\end{itemize}
Based on the considerations above, we categorize harnesses into two classes according to how they organize trajectories. Both participate in RL training but serve distinct roles: (1) \textbf{White-box harness}: mini-swe-agent. Its control flow is simple, it performs no trajectory compression, and its tool-invocation scale is relatively small. The trajectory structure is clear and closely mirrors the raw "task-solving" process. This type of harness provides clean, low-noise training signals, helping to directly strengthen the model's intrinsic, foundational agentic capabilities. (2) \textbf{Black-box harnesses}: ClaudeCode, Codex, OpenClaw, OpenHands and etc. These cover the diverse tool-exposure forms and control-flow complexities found in real deployment scenarios, and they commonly introduce trajectory-compression and context-reorganization mechanisms, making them the closest match to the actual deployment distribution. Training on this type of harness allows the model to adapt to incomplete-information conditions in which the context has been truncated, compressed, or reorganized, thereby improving its robustness and generalization in real-world complex environments.

\subsection{RL and Sandbox Infrastructure}

\begin{figure}[tbp]
\centering
\includegraphics[width=0.9\linewidth]{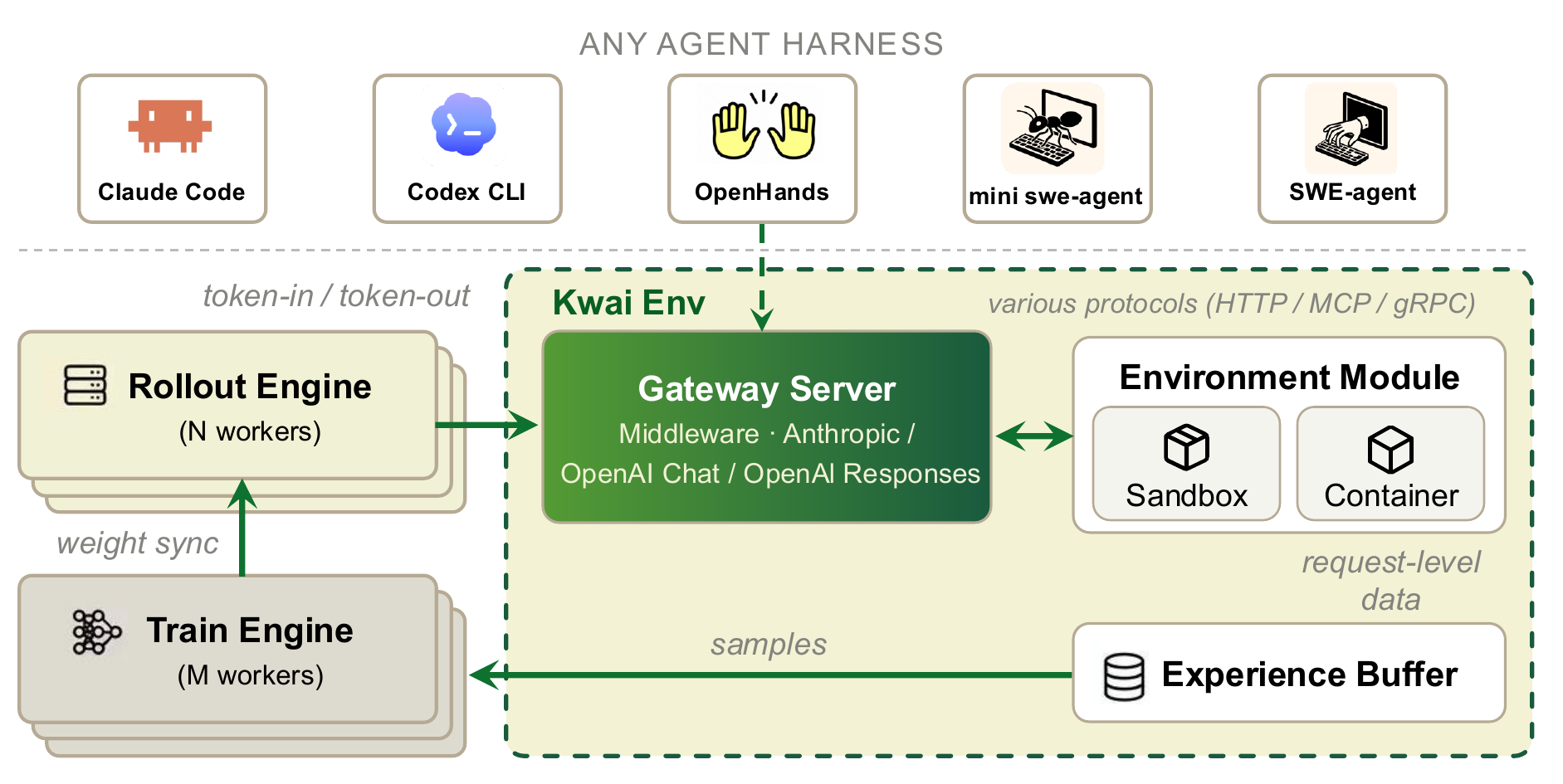}
\caption{Overall architecture of our Agentic RL training infrastructure.}
\label{fig:rl-infra}
\end{figure}

Similar to KAT-Coder-V2~\cite{zhan2025kat}, our training infrastructure consists of three core modules:
(1) \textbf{Rollout Engine}, which is responsible for policy-model inference and generates model responses during the agent's interaction with the environment;
(2) \textbf{Train Engine}, which performs policy updates based on the collected trajectories; and
(3) \textbf{KwaiEnv}, which manages environments and integrates harnesses by hosting corresponding execution environments.

\subsubsection{Gateway Server: Bridging the Model and the Harness}

To better support harness scaling, we introduce an additional \textbf{Gateway Server} module on top of KwaiEnv. The Gateway Server serves as an intermediary layer that connects the inference--interaction loop with the training loop. The Rollout Engine and the environment module do not communicate directly; instead, all traffic is mediated by the Gateway Server. Once a trajectory is completed, the Gateway Server writes it into the Experience Buffer, from which the Train Engine samples batches for policy updates.
This design decouples the inference--interaction loop from the training loop and fully isolates the execution details of the harness from the trainer. As a result, an arbitrary harness with an arbitrary execution environment can be mounted onto the training system as an opaque black box, without exposing its internal state machine to the trainer.

The second responsibility of the Gateway Server is to enforce token consistency. Mainstream inference engines typically expose a \texttt{chat} endpoint, which internally re-applies \texttt{apply\_chat\_template} and re-tokenizes the request. For long-horizon agentic tasks, the resulting token drift is non-negligible. In our experiments on agentic tasks at the $\sim$200-turn scale, we observe token drift in approximately \textbf{40\%} of the samples. This phenomenon, commonly referred to as \textbf{retokenization drift}, has also been documented in NVIDIA \href{https://arxiv.org/abs/2605.24220}{Polar}, as well as in the vLLM and \href{https://arxiv.org/abs/2508.03680}{Agent Lightning} ecosystems.
Our implementation bypasses the \texttt{chat} interface entirely and sends every request directly to the inference backend through its \texttt{/generate} endpoint. This eliminates retokenization drift at its source and guarantees that every trainable token is identical to the token emitted by the behavior policy during rollout.

\subsubsection{Sandbox Optimization: Improving Reward Signal Reliability}

During the early stages of agentic RL training for KAT-Coder-V2, we frequently encountered training crashes. More commonly, the reward curve improved only slowly, and convergence was substantially delayed. We initially attributed these failures to the RL algorithm itself, investing considerable effort in hyperparameter tuning and exploring new algorithmic variants. However, as we scaled up the number of training samples and introduced deeper supervision mechanisms, we found that a significant portion of training failures actually originated from the training environment rather than the learning algorithm. In a sampling-based audit of early rollouts, roughly $16\%$ of trajectories contained at least one failure attributable to the sandbox itself rather than to the model policy, which explains why tuning RL hyperparameters alone was insufficient to improve the reward curve.
These environment-related failures were often infrequent and difficult to reproduce, but their impact was severe. Once triggered, they could cause training samples to receive incorrect sandbox feedback, thereby corrupting reward signals and degrading the stability of RL training. In the most severe cases, a single sandbox boundary misalignment could cause the observations of the subsequent ${\sim}40$ rollout steps to become empty, corrupting the reward signal of the entire trajectory.

\begin{figure}[tbp]
\centering
\includegraphics[width=0.8\linewidth]{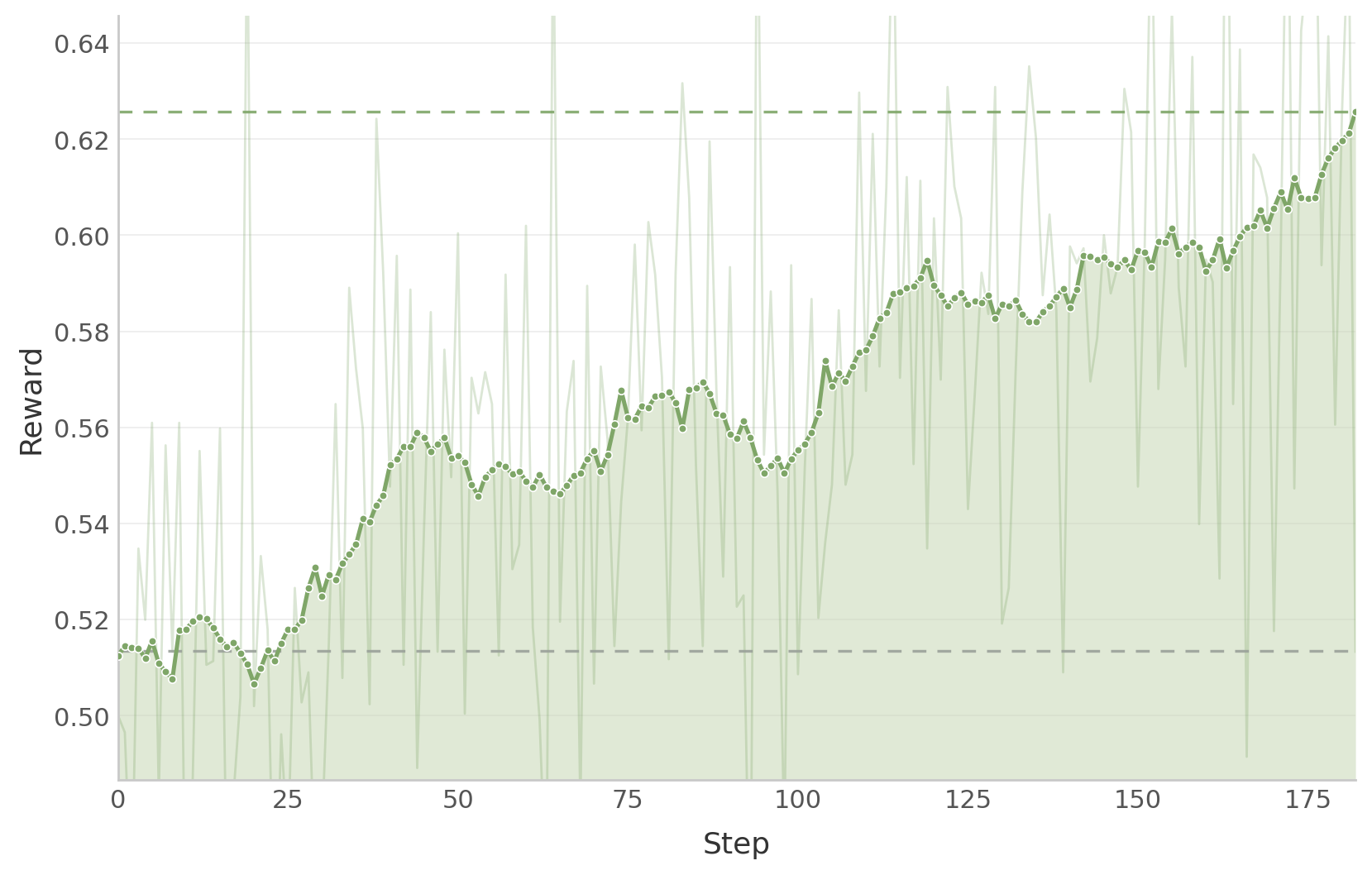}
\caption{RL training curve in the SWE scenario. The reward exhibits a consistently stable upward trend throughout training.}
\label{fig:rl-curve}
\end{figure}

During the development of KAT-Coder-V2.5, we devoted substantial effort to diagnosing and resolving a wide range of environment-related issues. These issues can be broadly categorized into two types: \textbf{sandbox stability} and \textbf{sandbox execution correctness}.

For sandbox stability, a representative example is container image management. RL training requires pulling a large number of container images concurrently, and the images used by our latest training environments are substantially large. This nearly exhausted the physical disk capacity of the machines hosting the sandbox service, which in turn triggered frequent garbage collection during execution. As a result, sandbox initialization and execution occasionally became slow enough to cause timeouts. Prior to optimization, peak-time disk usage reached roughly $95\%$, garbage collection ran almost continuously, and timeout-induced invalid rollouts accounted for approximately $6\%\text{--}7\%$ of all rollouts.

To address this issue, we redesigned the image management module of the sandbox service and introduced an early-release policy that proactively removes images unlikely to be reused in subsequent steps. After this optimization, steady-state disk usage dropped to roughly $60\%$, and the fraction of timeout-induced invalid rollouts fell to below $1\%$. This optimization also reduced sandbox cold-start latency and improved rollout efficiency during RL training.
Most sandbox execution correctness issues originated from bugs in our framework. For example, certain system environment variables set during remote sandbox initialization could override system configurations, causing some verifiers to read incorrect environment variables and produce erroneous verification results. Prior to the fix, this class of issues corrupted verifier outputs on roughly $6\%\text{--}7\%$ of samples, effectively flipping the corresponding reward signals. After the fix, the error rate dropped to below $1\%$.

After resolving the majority of these issues, training collapses became rare. Overall, the sandbox feedback error rate dropped from roughly $16\%$ to below $2\%$, and the frequency of training collapses decreased by approximately an order of magnitude. These improvements enabled the RL training of KAT-Coder-V2.5 to proceed smoothly and reliably.

\subsection{Asymmetric PPO for Long-Horizon Tasks}
\label{sec:rl-algo}

Agentic RL is naturally a long-horizon, partially observable, and stochastic-transition problem. In this setting, trajectory-level critic-free methods such as GRPO~\cite{shao2024deepseekmath} and its anchor-state variants~\cite{feng2026group} suffer from coarse credit assignment and high gradient variance. We therefore adopt Proximal Policy Optimization (PPO)~\cite{schulman2017proximal} as the algorithmic backbone.

PPO is preferable for three reasons. First, production-grade harnesses often produce multiple structurally split samples from the same session, e.g., after compacting, sub-agent splitting, or query rewriting. These samples may share the same final outcome but have different prefixes, making trajectory-level group baselines difficult to define consistently, whereas PPO supports per-token gradient contributions naturally. Second, retaining a Critic allows us to inject training-time privileged information, such as final rewards, test outcomes, and patch-level signals, into value estimation, reducing variance and improving sample efficiency. Third, PPO combined with Generalized Advantage Estimation (GAE) and reward shaping enables turn-level credit assignment, allowing us to penalize localized bad behaviors such as erroneous tool calls or regressive intermediate patches without uniformly affecting the whole trajectory.

\subsubsection{PPO Objective and Advantage Estimation}
\label{sec:rl-prelim}

We use the standard clipped PPO objective:
\begin{equation}
\label{eq:ppo}
\mathcal{J}_{\mathrm{PPO}}(\theta)
= \mathbb{E}_{q\sim P(Q),\ o\sim\pi'(\cdot\mid q)}
   \left[\frac{1}{|o|}\sum_{t=1}^{|o|}
   \min\!\left( r_t\,\hat{A}_t,\ 
   \mathrm{clip}\!\left(r_t,1-\epsilon,1+\epsilon\right)\hat{A}_t \right)\right],
\end{equation}
where $r_t=\pi_\theta(a_t\mid s_t)/\pi'(a_t\mid s_t)$ is the per-token
importance ratio and $\pi'$ is the behavior policy. The advantage is estimated using GAE:
\begin{equation}
\label{eq:gae}
\hat{A}_t
= \sum_{l=0}^{T-t-1}(\gamma\lambda)^l\delta_{t+l},
\qquad
\delta_t = r_t + \gamma V'(s_{t+1}) - V'(s_t),
\end{equation}
where $V'$ denotes the value network from the previous iteration. The Critic is trained by regressing to the return target $R_t=\hat{A}_t+V'(s_t)$:
\begin{equation}
\label{eq:critic}
\mathcal{L}_{\mathrm{critic}}(\psi)
= \mathbb{E}_{(s_t,R_t)\sim\mathcal{D}}
\left[\bigl(V(s_t;\psi)-R_t\bigr)^2\right].
\end{equation}

\subsubsection{Hindsight-Augmented Critic}
\label{sec:rl-asym-ac}

Following asymmetric actor-critic and centralized-critic paradigms~\cite{liu2025asymmetric,pinto2017asymmetric}, we let the Critic access privileged hindsight information during training, while the Actor only observes the normal harness state available at rollout time. This design improves value estimation without changing the deployed policy.

At rollout, the Actor observes only the interaction history up to the current turn, including tool outputs, file snippets, and compacted dialogue summaries. At training time, the Critic additionally receives a hindsight context $c_t$, which may include the final pass/fail reward, unit-test outcome distribution, coverage signals, patch-level differences, task metadata, trajectory statistics, and subsequent turns. The asymmetric value function is therefore written as $V_\psi(s_t,c_t)$.

The asymmetric Critic is trained with the same MSE objective:
\begin{equation}
\label{eq:critic-asym}
\mathcal{L}_{\mathrm{critic}}^{\mathrm{asym}}(\psi)
= \mathbb{E}_{(s_t,c_t,R_t)\sim\mathcal{D}}
\left[\bigl(V(s_t,c_t;\psi)-R_t\bigr)^2\right],
\qquad
R_t=\hat{A}_t+V'(s_t,c_t).
\end{equation}

GAE is computed as in Eq.~\ref{eq:gae}, with each value term replaced by the asymmetric value function $V(\cdot,c_t;\psi)$. The Actor still optimizes the standard PPO objective in Eq.~\ref{eq:ppo}. At inference time, the Critic and hindsight context are discarded, and only the Actor is deployed.

\subsection{Harness-Oriented Reward Framework}
\label{sec:rl-judge}


To address common challenges in harness-based reinforcement learning, including sparse reward signals, reward-hacking behaviors, non-standard operational patterns, and the lack of effective learning signals from failed trajectories, we design a harness-guided reward shaping framework that combines rule-based verification with model-based evaluation.

\subsubsection{Rule-based Reward}

We design a three-layer, structured rule-based reward system built on top of harness execution feedback, with the full set of rules summarized in Table~\ref{tab:rule-rewards}. The reward is organized into three tiers: \textbf{Core Task Score}, \textbf{Standard Behavior Constraints}, and \textbf{Failed Trajectory Incentives}. These tiers respectively capture final task success, process-level engineering discipline, and partial progress in incomplete trajectories. By jointly optimizing for task completion and standardized development behavior, the system provides fine-grained supervision for agent behavior.

\paragraph{Core Task Score.}
The Core Task Score is the primary reward component and carries the highest weight in the overall reward system. It serves as the main criterion for evaluating task completion and contains no sub-items.
A full score is awarded only when all \texttt{fail\_to\_pass} test cases, i.e., defect-revealing tests that should be fixed, and all \texttt{pass\_to\_pass} test cases, i.e., baseline regression tests that should remain stable, are passed. This design prevents the agent from exploiting reward-hacking strategies such as generating invalid code, bypassing intended logic, or weakening test cases. By requiring both bug-fixing success and regression stability, the Core Task Score anchors optimization to the true objective of the task: fully resolving the target defect while preserving existing functionality.

\paragraph{Standard Behavior Constraints.}

This layer applies auxiliary penalties throughout the entire interaction trajectory. Through rule-based verification, it penalizes inefficient, abnormal, or non-standard behaviors regardless of the final task outcome. The goal is to encourage standardized, efficient, and clean engineering behavior in long-horizon agent interactions. This layer includes the following components:
\begin{itemize}
  \item \textbf{Content Duplication Rate}: Penalizes redundant or repeated text in the thinking content and final response, reducing textual redundancy and improving information density.

  \item \textbf{Garbled Content Rate}: Penalizes garbled characters, illegal symbols, and other abnormal output patterns, reducing invalid content and avoiding downstream parsing errors.

  \item \textbf{Tool Call Accuracy}: Detects tool-call failures caused by missing or incorrect parameters, marked by \texttt{<tool\_use\_error>}, encouraging standardized tool invocation.

  \item \textbf{Abnormal Tool Call Position Rate}: Penalizes tool-call instructions incorrectly placed inside the \texttt{reasoning} field, ensuring that tool calls follow the expected format and can be parsed and executed reliably.

  \item \textbf{Repeated Single-Round Tool Calls}: Penalizes repeated calls to the same tool within a single interaction round, reducing redundant interactions.

  \item \textbf{Tool Call Parallelism}: Penalizes excessive parallel tool calls beyond a predefined threshold, while allowing reasonable batch calls to improve efficiency without compromising training stability.
\end{itemize}

\paragraph{Failed Trajectory Incentives.}

A key limitation of conventional reinforcement learning for agentic tasks is that incomplete or failed trajectories often receive zero reward. Consequently, the model obtains little feedback for partially correct intermediate behaviors, which weakens learning efficiency. To mitigate this issue, we introduce a failed trajectory incentive layer that assigns positive rewards to meaningful progress within failed trajectories, thereby densifying sparse reward signals. This layer contains two components:
\begin{itemize}
  \item \textbf{File Search Accuracy}: Evaluates repository file retrieval with an $F_2$ score that jointly considers precision and recall, encouraging the agent to locate relevant files while avoiding meaningless broad searches.

  \item \textbf{Unit Test Pass Rate}: Assigns rewards based on the pass status of the two types of unit tests. The model receives positive feedback when it passes any subset of test cases, providing useful learning signals for incomplete trajectories.
\end{itemize}

The reward tiers form a top-down, full-scenario guidance framework: (1) The top-level core reward anchors the ultimate task objective and prevents the model from deviating from the primary problem-solving goal. (2) The middle-layer behavior constraints uniformly regularize all trajectories, continuously improving the agent's output quality, tool usage, and engineering behavior. (3) The bottom-layer failed trajectory incentives compensate for reward sparsity and improve the training value of incomplete or failed samples.
Through the synergy of these reward components, the overall reward system improves the pass rate of code-repair tasks, reduces token consumption during interactions, and standardizes the behavior.

\subsubsection{Model-Based Reward}
\label{sec:model-based-reward}

Rule-based rewards are effective for easily verifiable signals, such as test outcomes and tool misuse, but they are insufficient for evaluating context-dependent behaviors. Assessing test sufficiency, strategy adaptation, and failure coverage requires holistic trajectory-level reasoning. Exhaustive hard-coded rules would introduce high complexity, poor generalization, and potentially unfair penalties for valid exploration. We therefore introduce a model-based judge, guided by a trajectory-level rubric, to complement rule-based rewards.

The rubric converts key code-repair behaviors into multi-dimensional process rewards, providing dense and fine-grained feedback beyond sparse task-level outcomes. It is designed to encourage robust, efficient, and auditable problem-solving behavior.

We construct the rubric through manual failure analysis of real sampled trajectories, covering tool sequences, environment feedback, code edits, and test operations. Table~\ref{tab:badcase} shows representative bad cases identified in this analysis. From these cases, we extract recurring, evidence-backed failure patterns that significantly impair task performance, and abstract them into standardized criteria with explicit triggers, applicable scopes, and reward signals.

Our analysis shows that performance bottlenecks often arise from unstable execution strategies rather than incorrect code edits alone. Typical failure patterns include running tests without validating entry points or environments, blindly scanning large files, overusing Bash instead of specialized code tools, and skipping targeted regression tests after fixes. We organize the model-based process rewards into three dimensions: \textbf{fault diagnosis and reproduction}, \textbf{post-fix verification}, and \textbf{execution strategy}. Representative examples are given in Table~\ref{tab:rubric-summary}.

\paragraph{GRM Training.}
The raw base model fails to strictly follow rubric rules during trajectory evaluation. We apply targeted RL to train a specialized judge model GRM for consistent rubric violation detection.
Training samples are historical agent trajectories paired with human annotations of triggered rubric criteria and rationales. We manually filter labels to keep only fully evidence-backed ones, eliminating ambiguous data to help the GRM learn reliable classification boundaries.
We define the trajectory reward for GRM optimization:

\begin{equation}
r =
\begin{cases}
\displaystyle \frac{|GT \cap Pred|}{|GT|} - \big|Pred \setminus GT\big| \cdot \lambda, & GT \neq \emptyset \\[6pt]
\displaystyle 1 - \big|Pred \setminus GT\big| \cdot \lambda, & GT = \emptyset
\end{cases}
\label{eq:grm-reward}
\end{equation}
The recall-based term maximizes ground-truth coverage, while the penalty term restrains excessive false predictions.
After RL, the GRM better incorporates full trajectory context when judging post-fix validation behaviors, reducing errors from superficial keyword matching.

\begin{table*}[t]
\centering
\caption{
Rule-based reward components. The rewards are organized into a primary task
objective, standard behavioral constraints, and failure-path incentives.
}
\label{tab:rule-rewards}

\small
\renewcommand{\arraystretch}{1.18}
\setlength{\tabcolsep}{5pt}

\resizebox{\textwidth}{!}{
\begin{tabular}{
>{\centering\arraybackslash}m{0.06\textwidth}
>{\centering\arraybackslash}m{0.25\textwidth}
>{\RaggedRight\arraybackslash}m{0.62\textwidth}
}
\toprule
\textbf{No.} &
\textbf{Reward} &
\centering\textbf{Description}\arraybackslash \\
\midrule

\multicolumn{3}{l}{\textit{Core Task Objective}} \\
\midrule

1 &
Core Task Score &
Uses the outcome of task-provided unit tests as the primary optimization
criterion. This reward prevents reward hacking through ineffective code changes
or simplified self-authored tests, keeping resolution of the underlying code
issue as the agent's highest-priority objective. \\
\midrule

\multicolumn{3}{l}{\textit{Standard Behavioral Constraints}} \\
\midrule

2 &
Content Repetition Rate &
Penalizes excessive redundant overlap between the \texttt{think} and
\texttt{content} fields, reducing unnecessary verbosity and improving output
quality. \\
\midrule

3 &
Garbled Content Rate &
Penalizes malformed, garbled, or illegal characters in generated text to reduce
low-quality output. \\
\midrule

4 &
Tool Invocation Accuracy &
Encourages valid and well-formed tool calls, reducing execution failures caused
by missing or incorrect parameters. \\
\midrule

5 &
Invalid Tool-call Placement &
Encourages the model to place tool calls in the designated output region rather
than in the \texttt{think} field, preventing parsing failures. \\
\midrule

6 &
Redundant Intra-turn Tool Calls &
Discourages repeated invocations of the same tool within a single interaction
turn, improving interaction efficiency. \\
\midrule

7 &
Tool-call Parallelism &
Encourages tool calls to be batched within an appropriate range, improving
interaction efficiency while avoiding excessive concurrency that may destabilize
training. \\
\midrule

8 &
Debug Artifact Cleanup &
Encourages removal of temporary reproduction scripts, validation files, logs,
and cache files after task completion, preventing repository pollution and
interference with subsequent execution or evaluation. \\
\midrule

\multicolumn{3}{l}{\textit{Failure-Path Incentives}} \\
\midrule

9 &
File Search Accuracy &
Measures whether the agent retrieves relevant repository files precisely and
completely. Retrieval quality is evaluated using an $F_{2}$ score that jointly
considers precision and recall, encouraging accurate file localization while
discouraging overly broad searches. \\
\midrule

10 &
Unit Test Pass Rate &
Provides partial learning signals for trajectories that do not completely solve
the task but make measurable progress without introducing regressions. This
reward distinguishes partially useful failures from entirely unproductive
trajectories. \\
\bottomrule

\end{tabular}
}
\end{table*}

\begin{table*}[t]
\centering
\caption{Representative trajectory-level rubric criteria. (Partial Display)}
\label{tab:rubric-summary}

\small
\renewcommand{\arraystretch}{1.18}
\setlength{\tabcolsep}{4pt}

\begin{tabular}{@{}
m{0.16\textwidth}
m{0.25\textwidth}
m{0.54\textwidth}
@{}}
\toprule
{\centering\textbf{Category}\par} &
{\centering\textbf{Representative Criterion}\par} &
{\centering\textbf{Evaluation Description}\par}
\tabularnewline
\midrule

\multirow[c]{2}{0.16\textwidth}[-1.0cm]{%
\centering\textbf{Bug Reproduction}\par
}
&
{\centering No Bug Reproduction Attempt\par} &
{\centering The task involves defect fixing, but the trajectory performs no
reproduction, test execution, or debugging step to validate the reported issue
before modification.\par}
\tabularnewline
\cmidrule(lr){2-3}

&
{\centering Static Bug Localization\par} &
{\centering The original failure is not dynamically reproduced, but the
trajectory accurately identifies a root cause consistent with the problem
statement through source inspection, existing tests, commit history, or
framework-level reasoning.\par}
\tabularnewline
\midrule

\multirow[c]{2}{0.16\textwidth}[-1.0cm]{%
\centering\textbf{Post-Fix Validation}\par
}
&
{\centering Missing Custom Validation\par} &
{\centering After applying the fix, the agent does not construct or execute a
targeted validation script for the failure scenarios explicitly described in the
problem statement.\par}
\tabularnewline
\cmidrule(lr){2-3}

&
{\centering Relevant Existing Tests Not Executed\par} &
{\centering Relevant repository tests exist for the reported failure scenario,
but the agent does not execute them after the fix. This criterion does not apply
when no relevant tests are available.\par}
\tabularnewline
\midrule

\multirow[c]{2}{0.16\textwidth}[-1.0cm]{%
\centering\textbf{Regression Testing}\par
}
&
{\centering Missing Broader Regression Testing\par} &
{\centering The agent validates only the directly affected failure scenario and
does not run broader tests to assess whether the modification introduces
regressions.\par}
\tabularnewline
\cmidrule(lr){2-3}

&
{\centering Invalid Broader Test Execution\par} &
{\centering The agent attempts broader testing, but the command or script fails
because of syntax, configuration, or invocation errors, leaving no valid
regression evidence.\par}
\tabularnewline
\midrule

\multirow[c]{2}{0.16\textwidth}[-1.0cm]{%
\centering\textbf{Behavioral Strategy}\par
}
&
{\centering Complex \texttt{python -c} Execution\par} &
{\centering The agent places lengthy or multi-step Python logic in an inline
\texttt{python -c} command, leading to escaping, encoding, readability,
debugging, or maintenance issues.\par}
\tabularnewline
\cmidrule(lr){2-3}

&
{\centering Repeatedly Incorrect Test Entry Points\par} &
{\centering The trajectory repeatedly invokes incorrect test commands, working
directories, test labels, frameworks, or module paths, resulting in avoidable
trial-and-error and unreliable validation.\par}
\tabularnewline
\bottomrule

\end{tabular}
\end{table*}

\begin{table*}[t]
\centering
\caption{Typical bad cases. (Partial Display)}
\label{tab:badcase}

\small
\renewcommand{\arraystretch}{1.18}
\setlength{\tabcolsep}{4pt}

\begin{tabular}{@{}
m{0.20\textwidth}
m{0.75\textwidth}
@{}}
\toprule
{\centering\textbf{Category}\par} &
{\centering\textbf{Example}\par}
\tabularnewline
\midrule

\multirow[c]{1}{0.20\textwidth}[0.15cm]{%
\centering\textbf{Complex code execution}\par
}
&
{\centering\texttt\{"function": \{ "name": "Bash", "arguments": "\{\textbackslash"command\textbackslash": "python -c \textbackslash"\textbackslash\textbackslash nfrom django.core.validators import URLValidator\textbackslash\textbackslash nfrom django.core.exceptions import ValidationError\textbackslash\textbackslash n\textbackslash\textbackslash nvalidator =......\par}
\tabularnewline
\midrule

\multirow[c]{1}{0.20\textwidth}[0.00cm]{%
\centering\textbf{Tool call failed}\par
}
&
{\centering\texttt<tool_response>\textbackslash n<tool_use_error>String to replace not found in file.\textbackslash nString:         return Perm(perm)\textbackslash n \textbackslash n    @classmethod......\par}
\tabularnewline
\midrule

\multirow[c]{1}{0.20\textwidth}[0.1cm]{%
\centering\textbf{Special tools not used}\par
}
&
{\centering\texttt \{"function": {"name": "Bash", "arguments": \{"command": "grep -r "DJANGO_SETTINGS_MODULE" /testbed/tests/*.py 2>/dev/null ...... \par}}
\tabularnewline
\midrule

\multirow[c]{1}{0.20\textwidth}[0.15cm]{%
\centering\textbf{Using non-existent modules}\par
}
&
{\centering\texttt \{"role": "tool", "tool_call_id": "call_53d347c9917c4b36bab2d659", "content": "/opt/miniconda3/envs/testbed/bin/python: No module named pytest"\}\par}
\tabularnewline
\bottomrule

\end{tabular}
\end{table*}

\section{Multi-Teacher On-Policy Distillation}
\label{sec:mopd}

After training domain-specific experts for agentic software engineering, general agentic reasoning, terminal use, web coding, and general knowledge, the key post-training challenge is to fuse their capabilities into a single unified student model. Conventional merging methods, such as parameter averaging, Task Arithmetic~\cite{ilharco2023task}, or sequential multi-domain SFT, often suffer from a \textit{see-saw effect}: improving one domain may degrade another. This is mainly because parameter-space merging can disrupt expert-specific structures, while offline data-space supervision introduces train--inference distribution mismatch.

To address this, we introduce \textbf{Multi-Teacher On-Policy Distillation} (MOPD), which performs capability fusion in the function space. Let the student policy be $\pi_\theta$, and the domain experts be $\{\pi_{T_k}\}_{k=1}^{K}$, where $K=5$ corresponds to five experts. For each sample $(x,d)$, where $d$ denotes its domain, the student first generates an on-policy trajectory $y\sim\pi_\theta(\cdot\mid x)$. The corresponding domain teacher $\pi_{T_d}$ then provides token-level logit supervision on the same trajectory. The student is optimized with reverse KL:
\begin{equation}
\label{eq:mopd_obj}
    \mathcal{L}_{\mathrm{MOPD}}(\theta)
    =
    \mathbb{E}_{(x,d)\sim \mathcal{D}}
    \mathbb{E}_{y\sim\pi_\theta(\cdot\mid x)}
    \left[
    \sum_{t=1}^{|y|} w_t\,
    \mathrm{KL}\!\left(
    \pi_\theta(\cdot\mid x,y_{<t})
    \,\middle\|\,
    \pi_{T_d}(\cdot\mid x,y_{<t})
    \right)
    \right],
\end{equation}
where $w_t\in[0,1]$ is a drift-aware token weight. The reverse KL objective is mode-seeking, encouraging the student to concentrate probability mass on the teacher's high-confidence regions and reducing cross-domain interference in multi-teacher distillation~\cite{agarwal2024gkd}.

\subsection{Stabilizing Long-Context On-Policy Distillation}
\label{subsec:mopd_stability}

Pure on-policy distillation becomes unstable on long-context tasks such as multi-turn agent trajectories and repository-level software engineering tasks. As the student-generated prefix grows, it may diverge from the teacher's training distribution. In this case, the teacher distribution conditioned on the student prefix becomes unreliable, producing biased token-level KL signals. This can lead to loss oscillation, entropy collapse, and gradient norm spikes, especially because reverse KL may further amplify over-confident collapse toward an incorrect local mode.
We address this issue with two stabilizing mechanisms: {off-policy cold start} and {drift-aware dynamic truncation}.

\paragraph{Off-policy cold start.}
Before on-policy MOPD, we initialize the student with expert-generated trajectories:
\begin{equation}
\label{eq:cold_start}
    \mathcal{L}_{\mathrm{cold}}(\theta)
    =
    \mathbb{E}_{(x,d)\sim \mathcal{D}}
    \mathbb{E}_{y\sim\pi_{T_d}(\cdot\mid x)}
    \left[
    -\sum_{t=1}^{|y|}
    \log \pi_\theta(y_t\mid x,y_{<t})
    \right].
\end{equation}
This phase pre-aligns the student with the teacher distributions, reduces early prefix divergence, and accelerates convergence in the subsequent on-policy stage.

\paragraph{Drift-aware dynamic truncation.}
Even after cold start, local student and teacher drift may still appear in the latter part of long trajectories. Following the top-$k$ overlap idea in Prune-OPD~\cite{yang2026pruneopd}, we define the compatibility between the teacher and student at token position $t$ as
\begin{equation}
\label{eq:compatibility}
    \rho_t =
    \frac{|\mathcal{T}_t^k \cap \mathcal{S}_t^k|}{k},
    \qquad
    \rho_t\in[0,1],
\end{equation}
where $\mathcal{T}_t^k$ and $\mathcal{S}_t^k$ are the teacher's and student's top-$k$ prediction sets. A high $\rho_t$ indicates that the teacher still provides reliable supervision under the current student prefix.

We use $\rho_t$ to control token weights and truncation. Specifically, $w_t$ in Eq.~\ref{eq:mopd_obj} is set according to a monotonic function of $\rho_t$, and tokens with low compatibility receive reduced or zero weight. If $\rho_t$ stays below a hard threshold for $m$ consecutive tokens, we truncate the trajectory and stop backpropagating through subsequent tokens. This removes unreliable teacher supervision while reallocating compute to new samples.
To avoid length bias caused by truncating long trajectories, we retain all valid prefix tokens before the truncation point, apply truncation only as gradient masking rather than an explicit optimization objective, and use length-stratified batching to preserve the proportion of long-context samples.
Together, MOPD combines multi-teacher on-policy reverse-KL supervision, off-policy cold start, and drift-aware truncation. This enables stable long-context distillation and effectively fuses all five domain capabilities—spanning agentic software engineering, general agentic reasoning, terminal use, web coding, and general knowledge—into a single unified student model without the severe cross-domain degradation typical of weight-space merging.

\section{Evaluation}
\label{sec:evaluation}
We conduct a comprehensive evaluation of KAT-Coder-V2.5 across the full spectrum of coding-agent competencies, spanning repository-level software engineering, long-horizon agentic tool use, and terminal and scientific coding. To ensure fair and reproducible comparison, all systems are evaluated under an identical protocol: unless otherwise noted, we adopt \textbf{Claude Code} as the unified evaluation harness and hold the tool set, context budget, execution environment, and decoding configuration fixed across models, so that the measured gaps are attributable to model capability rather than to scaffold or environment discrepancies.

\subsection{Benchmarks}
\paragraph{KAT Code Bench}
\label{sec:swe-katcodebench}

To measure the practical utility of coding agents in authentic software-engineering settings, we curate {KAT Code Bench}, a repository-level SWE suite distilled from Kuaishou's internal development workload. Unlike code completion, single-file repair, or competitive-programming tasks, KAT Code Bench evaluates the full engineering loop: parsing a natural-language requirement, navigating a real codebase to localize cross-file edit sites, following the repository's established design conventions, producing a minimal yet correct patch, and closing an executable verification loop that certifies the change introduces no regressions.
The suite is designed around four principles: \textbf{authenticity}, \textbf{verifiability}, \textbf{reproducibility}, and \textbf{discriminative power}. It spans 12 programming languages and covers heterogeneous frameworks, project structures, and change types, including defect fixes, feature completion, interface compatibility, behavioral-consistency correction, cross-module edits, and regression repair. Each task pins its base commit, runtime environment, and verification entry point, ensuring that all models operate under identical context, tooling, and resource budgets. This design isolates intrinsic SWE capability from variance introduced by scaffolds or environments.
During curation, we explicitly suppress recurrent sources of measurement noise, including irreproducible environments and flaky tests, verifiers that are over-coupled to the reference implementation and reject behaviorally correct but structurally different solutions, description--verifier mismatches, and over-templated tasks that leak the intended implementation path and weaken discriminative power. After iterative sampling, execution, human review, and trajectory auditing, we retain only tasks that stably reflect a model's true SWE capability.
Every task is solved by an agent inside a controlled sandbox. Beyond the final pass rate, we log the complete behavioral trajectory, making KAT Code Bench not only a cross-model benchmark but also a substrate for downstream data selection, failure attribution, and data augmentation.

\paragraph{KAT Claw Bench}

To evaluate how well LLMs perform in realistic business-oriented tool-use scenarios, we build {KAT Claw Bench}, a benchmark suite derived from Kuaishou's internal agent application needs. Existing Claw-type benchmarks exhibit significant limitations in practice: task granularity is too fine, mostly remaining at the script-level or single-point scope without assessing complete task chains; scenario coverage is insufficient for critical production needs such as project-level delivery, complex agent workflows, and business analysis; and evaluation tasks diverge from actual business contexts, failing to reflect real-world model performance. These gaps cause benchmark results to deviate from production reality, providing unreliable guidance for model selection and optimization. To address these limitations, {KAT Claw Bench} complements capability dimensions that general benchmarks cannot reach, more accurately reflects model usability and delivery value in real business contexts, and provides targeted evaluation evidence to support model selection and deployment decisions.

{KAT Claw Bench} is constructed through a structured pipeline to ensure authenticity, reproducibility, and evaluation reliability. We first collect and expand representative queries from real business scenarios, then filter them based on authenticity, feasibility, and evaluability. Retained queries are converted into complete benchmark tasks with standardized descriptions, required materials, category labels, difficulty annotations, and scoring rubrics. Scoring adopts automatic checks for objective outputs-such as file existence, numerical correctness, format compliance, and code executability-and hierarchical evaluation that combines automatic validation with human judgment for open-ended deliverables, with the specific combination determined by task type. To improve quality, each task undergoes multi-model review to verify self-containedness and scoring validity, as well as cross-model validation to detect ambiguous requirements or unstable criteria. Tasks with unclear instructions or inconsistent scoring behavior are revised or removed.

{KAT Claw Bench} is organized by an operation-oriented taxonomy covering seven major categories: personal productivity and office work, content creation and operations, software development and engineering, data analysis and insights, information retrieval and processing, automated monitoring and alerting, and investment analysis and decision-making. These tasks span different scenarios, including short video, live streaming, e-commerce, advertising, and workplace automation, with emphasis on realistic business workflows such as project-level software delivery, platform-aware content generation, decision-oriented data analysis, and continuous competitive intelligence tracking.

\subsection{Main Results}
\noindent Table~\ref{tab:three_category_bench} reports the head-to-head comparison of KAT-Coder-V2.5 against a panel of leading frontier models---GLM-5.1, GLM-5.2, Kimi-K2.6, and Opus 4.8---across all six benchmarks.

\paragraph{Repository-level software engineering.}
On the Coding category, KAT-Coder-V2.5 ranks second on both SWE-Bench Pro and KAT Code Bench, trailing only the frontier general-purpose Opus 4.8 while outperforming the GLM-5 series and Kimi-K2.6 by clear margins. Among all evaluated peers it is thus the strongest system short of the leading frontier model on realistic repository-level engineering, reflecting robust repository comprehension, cross-file localization, and verification-driven repair.

\paragraph{Long-horizon agentic tool use.}
On the Claw category, KAT-Coder-V2.5 attains the best overall result on PinchBench, edging out Opus 4.8 and surpassing the GLM-5 series by a wide margin. On the more demanding, business-grounded KAT Claw Bench it remains tightly competitive with the strongest proprietary and open peers. These results confirm that our agentic post-training, harness scaling, and large-scale multi-tool trajectory learning translate into robust workflow execution, tool selection, and state tracking under interactive environments.

\begin{table}[H]
\centering
\caption{
Comparison between KAT-Coder-V2.5 and open-source/proprietary models. The highest score for each benchmark is \textbf{bolded}, and the second highest is \underline{underlined}. PinchBench scores are averages (Avg) taken from \url{https://pinchbench.com/} (retrieved on July 2, 2026). Data points marked with * are for Kimi-K2.7-Code.
}
\label{tab:three_category_bench}

\begin{adjustbox}{max width=\textwidth}
\begin{tabular}{lccccc}
\toprule
 &
\makecell{KAT Coder\\V2.5} &
\makecell{GLM\\5.1} &
\makecell{GLM\\5.2} &
\makecell{Kimi\\K2.6} &
\makecell{Opus\\4.8} \\
\midrule

\multicolumn{6}{l}{\textit{Coding}} \\
SWE-Bench Pro      & \underline{65.2}  & 58.4 & 62.1 & 58.6 & \textbf{69.2} \\
KAT Code Bench     & \underline{53.1}  & 49.6 & 50.3 & 48.9 & \textbf{57.3} \\

\midrule
\multicolumn{6}{l}{\textit{Claw}} \\
PinchBench(Avg)         & \textbf{94.9}  & - & 87.0 &    $80.7^{*}$  & \underline{93.5} \\
KAT Claw Bench     & 85.5  & 84.4 & \underline{86.8} & 85.2 & \textbf{90.7} \\

\midrule
\multicolumn{6}{l}{\textit{AA Coding Index}} \\
Terminal-Bench 2.1 & 60.7 & 61.8 & \underline{77.9} & 73.0 & \textbf{84.6} \\
SciCode            & 50.3 & 43.8 & \underline{50.5} & \textbf{53.5} & \textbf{53.5} \\

\bottomrule
\end{tabular}
\end{adjustbox}
\end{table}

\paragraph{Terminal and scientific coding.}
On the AA Coding Index category, larger general-purpose models retain an edge on these broader, less repository-centric tasks; nevertheless KAT-Coder-V2.5 stays competitive on scientific programming, remaining on par with GLM-5.2. 

\noindent Taken together, the results delineate a clear capability profile: best-in-class agentic tool use and near-frontier repository-level software engineering, achieved with an optimization budget deliberately concentrated on the tasks that dominate real-world development.

\section{Conclusion}

In this report, we presented KAT-Coder-V2.5, a coding-focused agentic model built on the premise that the primary bottleneck to stronger coding agents lies in training infrastructure rather than model scale. We addressed this through an end-to-end agentic post-training framework: AutoBuilder reconstructs multilingual repositories into reproducible, executable environments with fail-to-pass and pass-to-pass verification, from which we regenerate self-contained tasks, recover near-miss trajectories, and distill supervision via process-aware filtering, while KwaiClawEnv synthesizes large-scale tool-use trajectories through a scalable closed-loop pipeline. On top of these environments, we scaled reinforcement learning with harness randomization, a reliability-hardened sandbox that cut the feedback error rate from roughly 16\% to below 2\%, an asymmetric actor--critic PPO with hindsight-augmented value estimation, and a harness-oriented reward framework, and finally fused five domain experts into a single student via Multi-Teacher On-Policy Distillation while mitigating the see-saw effect of weight-space merging.

Across six software-engineering and agentic benchmarks, KAT-Coder-V2.5 delivers the best agentic tool-use result on PinchBench and ranks second only to the frontier Opus 4.8 on repository-level software engineering, while remaining competitive on terminal and scientific coding. These results indicate that treating environment construction, trajectory quality, and RL stability as first-class systems problems yields a clear capability profile achieved by concentrating the optimization budget on the tasks that dominate real-world development. Extending verifiable environment construction to broader settings, strengthening long-horizon credit assignment, and improving generalization on terminal and scientific tasks remain promising directions, and we hope the methodology described here offers a reusable foundation for the next generation of autonomous coding agents.


\section{Contribution}
\setlength{\parindent}{0pt}
Contributors’ names are listed in alphabetical order by first name. 

\textbf{Core Contributors}

\begin{tabular}{@{}L{0.18\textwidth}@{\hspace{1em}}
                L{0.18\textwidth}@{\hspace{1em}}
                L{0.18\textwidth}@{\hspace{1em}}
                L{0.18\textwidth}@{\hspace{1em}}
                L{0.18\textwidth}@{}}
Bo Huang       & Fengxiang Li   & Hao Xu         & Haoyang Huang   & Hongyi Fu   \\
Jinhua Hao     & Kun Yuan       & Minglei Zhang  & Pengcheng Xu    & Shiyang Liu  \\
Wenhao Zhuang  & Yuze Shi       & Zongxian Feng  &                 &              \\
\end{tabular}

\textbf{Contributors}

\begin{tabular}{@{}L{0.18\textwidth}@{\hspace{1em}} L{0.18\textwidth}@{\hspace{1em}} L{0.18\textwidth}@{\hspace{1em}} L{0.18\textwidth}@{\hspace{1em}}
L{0.18\textwidth}@{}} 
Chao Wang & Cheng He & Chongling Rao & Deyu Cao & Fan Yang \\
Gang Xiong & Haochen Liu & Jiabao Li & Jian Liang & Jinghui Jia \\ 
Jingwen Chang & Jun Du & Junyu Shi & Min Li & Mingqi Wu \\ 
Qiang Gao & Shangpeng Yan & Shaotong Qi & Shu Xu & Shuo Zhou \\ 
Tiankuo Xu & Tong Zheng & Weilun Zhao & Xiancheng Meng & Xianda Sun \\ 
Xiaoyu Jiang & Xunhao Jia & Yao Xia & Yimeng Xu & Yinghan Cui \\ 
Yingpeng Chen & Yiwen Ning & Yong Wang & Yuxuan Sun & Zhongsheng Liu \\
\end{tabular}

\textbf{Tech Leads}

\begin{tabular}{@{}L{0.18\textwidth}@{\hspace{1em}}
                L{0.18\textwidth}@{\hspace{1em}}
                L{0.18\textwidth}@{\hspace{1em}}
                L{0.18\textwidth}@{\hspace{1em}}
                L{0.18\textwidth}@{}}
Ming Sun & Cheng Luo & Chen Yang & Han Li & Kun Gai \\
\end{tabular}

\newpage
\bibliography{references}

\end{document}